# MID-INFRARED ETHANE EMISSION ON NEPTUNE AND URANUS[1]

*Short Title: Ethane on Neptune and Uranus*

H. B. HAMMEL[2,*], D. K. LYNCH[3,*], R. W. RUSSELL[3,*],

M. L. SITKO[2,*], L. S. BERNSTEIN[4], and T. HEWAGAMA[5]

[1] Based on observations at the NASA Infrared Telescope Facility, operated by the University of Hawaii under contract to NASA.

[2] Space Science Institute, 4750 Walnut Street, Suite 205, Boulder, CO 80301. Electronic address for HBH: hbh@alum.mit.edu; electronic address for MLS: amazons@fuse.net

[3] The Aerospace Corporation, Mail Code M2-266, Los Angeles, CA 90009 Electronic address for DKL: David.K.Lynch@aero.org; electronic address for RWR: Ray.W.Russell@aero.org

[4] Spectral Sciences, Inc., 4 Fourth Avenue, Burlington, MA 01803. Electronic address: larry@spectral.com

[5] University of Maryland, Department of Astronomy, College Park, MD 20742; also at NASA's Goddard Space Flight Center, Greenbelt. Electronic address: tilak@cuzco.gsfc.nasa.gov

---

* *Visiting Astronomer at the Infrared Telescope Facility, which is operated by the University of Hawaii under Cooperative Agreement no. NCC 5-538 with the Planetary Astronomy Program of NASA's Science Mission Directorate.*

ABSTRACT

We report 8- to 13-$\mu$m spectral observations of Neptune and Uranus from the NASA Infrared Telescope Facility spanning more than a decade. The spectroscopic data indicate a steady increase in Neptune's mean atmospheric 12–$\mu$m ethane emission from 1985 to 2003, followed by a slight decrease in 2004. The simplest explanation for the intensity variation is an increase in stratospheric effective temperature from 155 ± 3 K in 1985 to 176 ± 3 K in 2003 (an average rate of 1.2 K/year), and subsequent decrease to 165 ± 3 K in 2004. We also detected variation of the overall spectral structure of the ethane band, specifically an apparent absorption structure in the central portion of the band; this structure arises from coarse spectral sampling coupled with a non-uniform response function within the detector elements. We also report a probable direct detection of ethane emission on Uranus. The deduced peak mole fraction is approximately an order of magnitude higher than previous upper limits for Uranus. The model fit suggests an effective temperature of 114 ± 3 K for the globally-averaged stratosphere of Uranus, which is consistent with recent measurements indicative of seasonal variation.

*Subject Headings:* planets and satellites: Uranus, Neptune -- infrared: solar system.

## 1. INTRODUCTION

In the atmospheres of Uranus and Neptune, the most abundant constituent after hydrogen and helium is methane ($CH_4$). Photolysis of $CH_4$ at pressures near a microbar produces ethane ($C_2H_6$), and the abundance of this molecule as a function of altitude is an important tracer of atmospheric microphysical properties. The presence of $C_2H_6$ in the atmosphere of Neptune, and its marked dearth in the atmosphere of Uranus, was first noted more than a quarter century ago (Gillett & Rieke 1977).

In 2003, Neptune reached its brightest level in nearly 30 years of photometric monitoring (Lockwood & Jerzykiewicz 2006; Hammel & Lockwood 2006). The planet (Fig. 1) has exhibited pronounced cloud activity throughout this time (Max et al. 2003; Sromovsky et al. 2003), possibly undergoing a transformation from one cloud distribution to a new configuration (Hammel & Lockwood 2006). Hammel & Lockwood (2006) predicted that the stratospheric temperature may have increased on Neptune since the mid 1980s, because stratospheric temperatures on Uranus (Young et al. 2001) rose and fell in lockstep with that planet's visible wavelength reflectivity (Hammel & Lockwood 2006). Mid-infrared methane- and ethane-band emissions are sensitive to temperature, thus a measurement of Neptune's hydrocarbon emission levels in the current epoch could test this hypothesis.

Uranus reaches equinox in 2007. As the equatorial and northern regions of Uranus experience increasing insolation, the planet's appearance (Fig. 2) has altered due both to the changing viewing angle, as well as to intrinsic physical change (Hofstadter & Butler 2003; Hammel et al. 2005a, 2005b). Such images have led to speculation that the increasing number of cloud features are associated with enhanced production of hydrocarbons (J. Moses, personal communication) and with more atmospheric dynamic activity in general (Hammel et al. 2005b). One manifestation of enhanced dynamics could be increased generation of upwardly propagating gravity waves and hence an increased eddy diffusion coefficient in the stratosphere. Methane would get carried up to higher altitudes, and photochemical products like acetylene and ethane would be produced at a higher rate and mixed into the stratosphere. Thus, a consequence of enhanced eddy mixing in Uranus would be the ability to detect trace hydrocarbon species more readily than before. It is intriguing that Encrenaz et al. (1998), using 1996 observations with the Infrared Space Observatory (ISO), derived an acetylene ($C_2H_2$) abundance on Uranus that is 30 times greater than was seen during the 1986 Voyager encounter (Bishop et al. 1990). Encrenaz et al. (1998) set a higher upper limit on the ethane mixing ratio of 3 x$10^{-6}$, again using ISO.

Motivated by the discernable changes in the visible and near-IR appearances of Neptune and Uranus, and by the suggestion that those changes may be linked to photochemistry, we set out to remeasure the $C_2H_6$ abundances on these planets beginning in 2002, and to compare the result with the historical observations. For Neptune, confirmation of increased temperatures could provide insight into the photochemical pathways that link the stratosphere and troposphere. For Uranus, a measurable ethane signal would constitute a significant milestone in understanding the atmosphere of Uranus, and would provide the basis for future observations to pursue variability in stratospheric chemistry and dynamics.

## 2. OBSERVATIONS

We obtained infrared spectroscopy of Neptune, Uranus, and calibration stars using NASA's Infrared Telescope Facility (IRTF) 3-m telescope on Mauna Kea in 2002, 2003 and 2004 with the Broadband Array Spectrograph System (BASS; Hackwell *et al.* 1990); the observations are tabulated in Table 1. BASS records the spectra on two 58-element back-illuminated blocked-impurity-band (BIBIB) arsenic-doped silicon arrays, one for shorter wavelengths (3 to 6.5 $\mu$m) and the other for longer wavelengths (6.5 to 13.5 $\mu$m). This paper presents data from the long-wavelength array only, which has a resolution of 0.1 to 0.25 $\mu$m (~2%). The nominal BASS wavelengths were calibrated with a chopped monochromer.

We also calculated additional flux corrections for the finite angular size of Neptune (Fig. 1), since more light was scattered out of our 3.2" aperture for Neptune's 2.3" disk than for a point source (we used 24760 km for Neptune's radius). However, this correction was small, much less than the intrinsic variation of the planet (discussed below). The 3.4" disk of Uranus (Fig. 2) was larger than the aperture, thus this correction was not applied.

## 3. NEPTUNE

Gillett & Rieke (1977) first detected ethane and methane emission in Neptune's mid-infrared spectral region at 2% resolution in 1975 and 1976. Measurements were made again in 1977 (Macy & Sinton 1977) and in 1981 and 1982 (Orton *et al.* 1983); the data were consistent with the results of Gillett & Rieke (1977) to within observational errors. In 1985, 1986, and 1987, Orton *et al.* (1987, 1990, 1992) re-observed Neptune at 8-14 $\mu$m. They reported that the spectral radiance of the 12.2-$\mu$m ethane feature was possibly 25% higher than that seen ten years earlier. However, the change in emission could not be assigned statistical significance because of internal uncertainties and the uncertainties in absolute levels: 30% and 20% for Gillett & Rieke (1977) and Orton *et al.* (1987, 1990) respectively. Nevertheless, the aggregate data raised the possibility that time-dependent variations occurred on Neptune's mid-infrared emission. Using independent observations in 1991, Hammel *et al.* (1992) and Kostiuk et al. (1992) confirmed the increase in ethane emission; Hammel et al. also reported statistically significant variability in the 12.2-$\mu$m ethane emission from night to night.

Figure 3 shows our 2002-2004 spectra resulting from combining all Neptune integrations in each year. We also show yearly-averaged spectra taken in 1991 with an earlier version of BASS (Hammel et al. 1992), and in 1985 with different instrumentation at the IRTF (Orton et al. 1990, 1992). The ethane emission increased steadily from 1985 to 2003, before growing dimmer between 2003 and 2004. As discussed below, the data also show significant variation in shape.

### 3.1 Model Description

In order to investigate potential sources of the observed variability in the ethane emission spectrum, we first consider a single-layer, line-by-line radiative transfer (RT) model tied to the atmospheric profiles shown in Fig. 4. As in earlier similar models (e.g., Orton et al. 1992; Moses et al. 1992), the ethane molecular emission arises from hot ethane in the stratosphere, i.e., predominantly from 1 to 0.1 mbars. Orton et al. (1992) obtained good model agreement with moderate- and high-resolution observations in the 1980s and 1990s using a constant ethane mixing ratio of $1.5 \times 10^{-6}$ at pressures below 10 mbar as illustrated in Fig. 4b. (A number of profiles, corresponding to temperature variations about this profile, yielded similar spectra, but this model profile was slightly favored and we adopt it as the reference profile for our study.) Fig. 4b also shows a different ethane mixing ratio profile based on photochemical modeling (Moses et al. 1992); the Moses et al. profile was scaled by an altitude-independent constant to yield the same spectral emission as the nearly altitude-independent profile of Orton et al. (1992). Our ethane line parameters were obtained from the same source used in the Orton work (W. Blass 2005, private communication); the acetylene parameters were obtained from the HITRAN database (Rothman et al. 2003).

Since the temperatures and mixing ratios based on Orton's profiles are nearly constant over the atmosphere from 0.1 to 1 mbar, it was anticipated that a single-layer approach using a scale-height-based ethane column amount would work well (scale-height = 20 km). We verified this via comparisons to Orton's moderate and high-resolution spectral predictions using temperature T = 155 K, ethane mixing ratio $\chi = 1.5 \times 10^{6}$, pressure P = $1.6 \times 10^{-3}$ bar, and ethane column amount $u$ = $1.7 \times 10^{-3}$ atm-cm (at standard temperature and press) to represent Orton's profiles. Acetylene is responsible for most of the enhanced flux longward of 12.9 $\mu$m. We reference the acetylene spectra to the same profile but with an acetylene mixing ratio of $3.8 \times 10^{-8}$, $1/20^{th}$ of the ethane mixing ratio, as determined by Orton et al. (1992) for the profile with T = 160 K. The ethane column amount used in the RT calculations is a factor of two higher than the vertical column amount and represents the mean line-of-sight column amount averaged over the planetary disk.

As shown previously by Orton et al. (1992), there is a non-unique dependence of the overall intensities of the ethane and acetylene emissions on both their mixing ratios and emission temperatures. For simplicity, in our analysis we fixed the mixing ratios at the reference values discussed above and retrieved an effective temperature, $T_{eff}$, to describe the year-to-year variations of the ethane and acetylene intensities.

First, however, we discuss in detail the absorption notch that appears near the peak of the 12-$\mu$m ethane emission (Fig. 3). We investigated whether instrumental effects could give rise to this observed structure. Through laboratory testing, we determined that the detector elements appeared to have a non-uniform response, likely due to fringing effects within the detector substrate. We also noted from precise astrophysical calibrations that very slight changes in wavelength sometimes occurred within the detector. These normally are undetectable, being only fractions of a detector-element width. However, Fig. 5 illustrates that the ethane spectrum of Neptune is rich and complex in the 12-micron region. Even miniscule shifts of wavelength (solid vs. dashed curves in Fig. 5) can result in the integrated flux within a detector element varying at a detectable level (inset in Fig. 5). Effectively, shifting the wavelength sensitivity of the detector element just a tiny amount slides specific ethane lines across the structure in the detector-element response function. For a single isolated line, one might not be able to distinguish such an effect from variations in terrestrial transparency (for example) or other such effects. However, Neptune's ethane band is comprised of many ethane lines spread across many BASS detector elements, and the resulting spectra are thus sensitive to the non-uniformity of the response function. Our models of this sensitivity yielded a variety of apparent band structures (Fig. 6), which match the observed structures to within the uncertainties.

### 3.2 Results from Neptune models

Figure 7 plots our data and several models that explore the sensitivity of ethane and acetylene emission to variations in the key model parameters. The model results are non-unique; the models shown here (Table 2) are representative. Nevertheless, some overall conclusions can be drawn. The amplitude of the 12-micron emission is dominated by effective temperature ($T_{eff}$; see above). A steady rise in temperature occurred on Neptune from 1985 to 2003, with $T_{eff}$ = 155, 160, 170, and 176 K in 1985, 1991, 2002, and 2003 respectively. In 2004, $T_{eff}$ dropped to 165 K. These temperatures have a representative range of ±3 K; i.e., temperatures outside that range exhibit amplitudes inconsistent with the observations.

We reiterate that because our data are disk integrated, we cannot distinguish a moderate-amplitude planet-wide change from a larger-amplitude change in one or more local atmospheric regions. $T_{eff}$ represents a global average of differing temperatures and mixing ratios. Either the overall background temperature rose, or a fraction of the disk with higher temperatures and/or mixing ratios increased, or perhaps some combination of the two occurred. For example, by fitting the Neptune data specifically for ethane column abundance, we find a vertical column of molecular ethane $u \geq 0.2$ atm-cm, compared with ~0.1 atm-cm for Orton's profile (see Fig. 4) and the theoretical vertical ice column predicted by Romani *et al.* (1993) of ~0.01 atm-cm. In short, we cannot tightly and uniquely constrain the underlying model parameters. We have ascribed the observed intensity variations to temperature variations from 155 K to 176 K for a fixed mixing ratio of $1.5 \times 10^{-6}$. Equivalently, we can ascribe the changes to mixing ratio variations from $0.6 \times 10^{-6}$ to $1.5 \times 10^{-6}$ for a fixed temperature of 155K. Below, we discuss independent observations that suggest the temperature variations are more probable. Furthermore, the ethane contribution function peaks in a region that is roughly isothermal (Fig. 4c), thus variations of the ethane abundance will not have much effect. Direct imaging of Neptune at wavelengths sensitive to the molecules of interest, or spectra with significantly higher resolution, should help discriminate among possible scenarios.

These disk-integrated mid-infrared spectra alone are inadequate to determine the precise nature of the change of Neptune's atmosphere, but results from other observational programs can provide additional context. The long-term photometric program at Lowell (Lockwood & Jerzykiewicz 2006) has shown a steady rise in disk-integrated visible brightness (Fig. 8). Sromovsky et al. (2003) attributed the rise to simple seasonal variation, but extensive reanalysis of earlier photometry showed this cannot explain all the observations (Lockwood & Jerzykiewicz 2006; Hammel & Lockwood 2006). Disk-resolved imaging programs from HST, Keck, and other facilities indicate that the amount of discrete cloud structure at visible and near-infrared wavelengths (e.g., Fig. 1) was in general lower in the mid- to late-1980s, when the first two sets of mid-IR data were obtained. However, the highest-sensing filters for those images programs reach only the lower stratosphere, well below the level from which the ethane emission is coming. If a mechanism could be identified that dynamically linked the troposphere and stratosphere, this may elucidate the rising temperatures and brightnesses.

Recent sub-mm observation from the James Clark Maxwell Telescope (Marten et al. 2005) are of particular interest, since some CO emission arises from roughly the same level in the atmosphere as the BASS ethane emission. Data obtained in September 1998 suggested an increase of 12 K in the stratosphere (pressures < 0.1 mbar) over nominal models based on the 1989 Voyager-era data (e.g., Bishop et al. 1998), and an increase of 8 K over a nominal model used for modeling ISO data taken in late 1997 (Bézard et al. 1999). These results, plotted in Fig. 8 along with the BASS effective temperatures, are consistent with the temperature increases we deduce from the BASS observations. Stratospheric temperatures inferred from occultation observations through 1990 are consistent with our early measurements, but show a broad range of variations (see summary by Roques et al. 1994). Their observed dispersion may be due to horizontal and/or vertical inhomogeneity, because occultations probe a specific latitude, longitude, and altitude in the planet's atmosphere, whereas the mid-IR and sub-mm data are usually averaging over the entire disk of the planet.

Recent disk-resolved 12-$\mu$m images obtained with the Gemini 8-m telescope (Hammel et al., in preparation) show the bulk of Neptune's 12-$\mu$m emission arising from two regions, a bright south polar region and bright emission distributed nearly uniformly around the planet's limb. Because the bulk of the non-limb emission is confined to the pole, seasonal variation may play some role in the increasing mid-infrared brightness: more of that region was revealed as Neptune's solstice neared in 2005. However, the 12-$\mu$m emission showed a significant drop in 2004 prior to solstice (Figs. 7 and 8). Thus intrinsic atmospheric variability is still required to fully explain our observations.

If the short-term variations in ethane emission are attributed to changes in atmospheric temperature, is this consistent with a purely radiative response? Probably not: the radiative time constant for Neptune's stratosphere is nearly two decades (Conrath et al. 1992), thus a 10-K drop between 2003 and 2004 (Fig. 8) is unlikely due to radiative effects. Our observations—and interpretation thereof—are limited due to the data being both disk-integrated and of moderate spectral resolution. Observations with higher spatial and spectral resolution have already been obtained, and may provide insight on the mechanisms underlying the observed changes (Hammel et al., in preparation).

## 4. URANUS RESULTS

In the mid 1980s, Orton *et al.* (1983) reported 12-$\mu$m observations of Uranus with the IRTF that yielded an ethane mixing ratio upper limit of $3 \times 10^{-8}$, later revised to an upper limit of $2 \times 10^{-8}$ (Orton *et al.* 1987). Other upper limits for ethane mole fractions for Uranus range from $2 \times 10^{-8}$ to $4 \times 10^{-8}$ both from observations (Herbert et al. 1987) and from photochemical model predictions (Fegley et al. 1991). In our Uranus spectra, the signal from the planet is extremely weak; thus all the data were combined to produce a single spectrum (Fig. 9). An intriguing hint of enhanced emission appears in the 12-$\mu$m region where ethane would be expected. The signal-to-noise ratio is low, and thus the detection is admittedly marginal. However, the analysis below supports the notion that the detection is real.

To model the Uranus spectrum, we begin with the atmospheric profiles shown in Fig. 4 (P. Romani, personal communication). To first order, ethane emission on Uranus is characterized by a lower effective black body temperature (~110 K; c.f. Neptune's ~155 K in 1985) and a smaller vertical column amount (~$6.4 \times 10^{-4}$ atm-cm at STP; c.f. ~$8.5 \times 10^{-3}$ atm-cm for Neptune). The ethane emission should scale approximately as the product of the Planck function and column amount, thus Neptune's emission is expected to be ~500 times larger than that of Uranus, i.e., the 1985 Neptune data would indicate ~$3 \times 10^{-10}$ W/sr/cm$^2$/μm for the 12.2-μm peak of the Uranus spectrum, consistent with our Uranus data which provides an upper limit of ~$6 \times 10^{-10}$ W/sr/cm$^2$/μm for the ethane peak (Fig. 9). The peak value for the ethane mixing ratio in the Uranus profile used here is $4 \times 10^{-7}$, an order of magnitude larger than the previous upper limit estimates. This higher value appears consistent with our observations, as shown in the more detailed spectral modeling below.

The specific spectral model for the Uranus data consisted of three additive components: ethane emission, acetylene emission, and H$_2$ continuum emission. We computed ethane and acetylene contributions using a line-by-line approach based on a ten-layer representation of the reference profile over the $10$-$10^{-2}$ mbar region. In the spectral fitting, we adjusted the reference temperature profile by an additive constant, and applied this to both the ethane and acetylene model emissions. The acetylene mixing ratio profile was modeled via a multiplicative scale factor applied to the ethane profile, and the scaling factor was adjusted in the spectral fit.

We initially calculated the H$_2$ continuum emission with a multi-layer representation of the reference atmospheric profile using the Borysow et al. (1985) absorption coefficients. However, the resulting spectral profile for Uranus could be very closely approximated as a black body curve. To simplify the calculations, we modeled the H$_2$ as a black body with an effective temperature determined by the fitting. (Collision-induced H$_2$ emission continua for both Uranus and Neptune arise primarily from the tropopause, ~$10$-$10^3$ mbar, at about the same effective temperature, ~55K, and therefore are of comparable levels; however, it was not necessary to include the H$_2$ component for Neptune because the much larger ethane emission completely dominates the spectral window).

Fig. 9 shows a representative spectral fit compared to the Uranus data. The fit parameters indicate an effective temperature of 114 K ± 3 K (slightly warmer than the reference profile of 110 K), an effective H$_2$ continuum temperature of ~62 K (about ~7 K higher than expected from the reference profile), and a scaling factor for the acetylene mixing ratio profile of ~0.2 relative to the ethane profile. Given the large error bars on these data, these fits are non-unique. Nevertheless, they suggest a slight change in temperature of the stratosphere on Uranus. Temperatures inferred from occultation observations (analyzed by Young et al. 2001) show an apparent stratospheric warming and cooling correlated with season, and seasonal effects are also seen at tropospheric levels (Fig. 2). The effective temperature reported here us consistent with this trend (Fig. 10).

Shortward of 11.5 μm in the Uranus spectrum, some other species (even though two orders of magnitude less in mixing ratio than ethane) could be contributing flux. However, the error bars are large and the data at the noise level. Spitzer observations may provide insight on this wavelength region, and thus we prefer to defer quantitative speculation based on these data.

## 5. CONCLUSIONS

In our mid-infrared spectra of Neptune at 7 to 13 μm, ethane emission centered at 12.2 μm markedly increased over the past decade, from 1985 until 2003, followed by a slight drop in 2004. This may be correlated with the steady increase of brightness seen at visible wavelengths. Such a correlation could indicate that stratospheric temperature variations have affected hydrocarbon creation at upper

Ethane on Neptune and Uranus					8altitudes, and would have implications for the dynamical and transport properties of the upper troposphere and lower stratosphere. We have obtained coordinated mid-infrared spectroscopy and adaptive-optics near-infrared imaging with Gemini and Keck in July 2005 to explore the potential correlation; the results from those data are in preparation.

We report a probable detection of blended ethane, acetylene, and $H_2$ continuum emission in the mid-infrared spectrum of Uranus, along with slight changes in stratospheric temperature that are consistent with seasonal change. As that planet continues its inexorable march toward equinox in 2007, additional observations are warranted to confirm this detection and to look for evidence of increased hydrocarbons as the planet's overall activity increases.

ACKNOWLEDGEMENTS

HBH acknowledges support for this work from NASA grants NAG5-11961 and NAG5-10451. This work was supported at The Aerospace Corporation by the Independent Research and Development Program. LSB acknowledges the support of Spectral Sciences, Inc. (SSI) IR&D funding. TH acknowledges the support of cooperative agreement NCC 5-609 with NASA's Goddard Space Flight Center. We acknowledge technical discussions with J. Pearl (NASA GSFC) on the properties of planetary ices, with W. Blass (University of Tennessee) on ethane line properties, and with L. Berk (SSI) on radiative-transfer issues. We thank P. Romani (NASA GSFC) for atmospheric profiles. We also gratefully acknowledge the support of the staff and telescope operators at the NASA IRTF, and D. Kim (The Aerospace Corporation) for BASS technical support. We recognize the significant cultural role of Mauna Kea within the indigenous Hawaiian community, and we appreciate the opportunity to conduct observations from this revered site.

REFERENCES

Baines, K. H., & Hammel, H. B. 1994, Icarus, 109, 20
Baines, K. H., et al., 1995, in "*Neptune and Triton*," U. Ariz. Press
Bézard, B., et al. 1999, ApJ, 515, 868
Bishop, J., et al. 1990, Icarus, 88, 448
Bishop, J., et al. 1998, Planet. Space Sci., 46, 1
Borysow, J., et al. 1985, ApJ, 296, 644
Conrath, B., et al. 1992, Icarus, 83, 255
de Pater, I., et al. 2005, Icarus, in press
Encrenaz, T., et al. 1998, A&A, 333, L43
Fegley, B., et al. 1991, in "*Uranus*," U. Ariz. Press
Gillett, F. C., & Rieke, G. H. 1977, ApJ, 218, L141
Hackwell J. A., et al. 1990, SPIE, 1235, 171
Hammel, H. B., & Lockwood, G. W. 2006, Icarus, submitted
Hammel, H. B., et al. 1992, Icarus, 99, 347
Hammel, H. B., et al. 1995, Science 268, 1740
Hammel, H. B., et al. 2005a, Icarus, 175, 284
Hammel, H. B., et al. 2005b, Icarus, 175, 534
Herbert, F., et al. 1987. JGR, 92, 15093
Hofstadter, M. D., & Butler, B. J. 2003, Icarus 165, 168
Kostiuk T., et al. 1992, Icarus, 99, 353
Lockwood, G. W., & Jerzykiewicz, M. 2006, Icarus, in press
Macy, W., & Sinton, W. 1977, ApJ, 218, L79
Marten, A., et al. 2005, A&A, 429, 1097
Max, C. E., et al. 2003, AJ, 125, 364
Moses, J. A., et al. 1992, Icarus, 99, 318
Orton, G. S., et al. 1983, Icarus, 56, 147
Orton, G. S., et al. 1987, Icarus 70, 1
Orton, G. S., et al. 1990, Icarus, 85, 257
Orton, G. S., et al. 1992, Icarus, 100, 541
Pearl, J., et al. 1991, JGR, 96, 17477
Romani, P., et al. 1993, Icarus, 106, 442.
Roques, F., et al. 1994, A&A, 288, 985
Rothman, L. S., et al. 2003, JQSRT 82, 5
Sromovsky, L. A., et al. 2003, Icarus, 163, 256
Young, L. A., et al. (2001), Icarus 153, 236

TABLE 1

URANUS AND NEPTUNE MID-IR OBSERVATIONS FROM 2002 TO 2004

| UT Date | Neptune* | Uranus† |
|---|---|---|
| 2002 Jul 26 | 2 | 10 |
| 2002 Jul 27 | 4 | 2 |
| 2002 Jul 28 | 7 | 6 |
| 2002 Jul 29 | 10 | 14 |
| 2002 Jul 30 | 9 | 14 |
| 2002 Jul 31 | 8 | 4 |
| 2002 Aug 1 | 6 | 14 |
| *2002 Total* | *46 sets = 2.56 hrs* | *64 sets = 7.11 hrs* |
| 2003 Aug 20 | 6 | - |
| 2003 Aug 21 | 15 | - |
| *2003 Total* | *21 sets = 1.17 hrs* | - |
| 2004 Aug 4 | 3 | - |
| 2004 Aug 5 | 10 | - |
| 2004 Aug 6 | 13 | - |
| *2004 Total* | *26 sets = 1.44 hrs* | - |

* Neptune: Number of 200-second sets.
† Uranus: Number of 400-second sets.

TABLE 2

REPRESENTATIVE NEPTUNE MODELS *

| Year | $T_{eff}$(K) | $\Delta\lambda$ ($\mu$m) |
|---|---|---|
| 1995 | 155 ± 3 | +0.074 |
| 1991 | 160 ± 3 | +0.040 |
| 2002 | 171 ± 3 | -0.030 |
| 2003 | 176 ± 3 | -0.100 |
| 2004 | 165 ± 3 | -0.010 |

* The temperatures correspond to the Neptune models shown in Fig. 7. The uncertainties are not formal errors, but rather estimates based on the dispersion of the data with respect to the model. The wavelength shifts were applied to the Neptune spectra in Figs. 3 and 7, and the Uranus spectra in Fig. 9, and are consistent with uncertainties in the calibration of the data.

# FIGURE CAPTIONS

Fig. 1.—Representative Neptune images during the past decade. These images of Neptune's 2.3" disk—taken with the Hubble Space Telescope in the WFPC2 619-nm methane-band filter and displayed with identical stretches—were obtained in 1994, 1997, 2001, and 2004 (from left to right, respectively). The 1994 image of Neptune shows the bright companion to the northern Great Dark Spot; this anomalously bright feature faded within months (Hammel et al. 1995). Subsequent images show southern mid-latitude activity increasing to a maximum in 2003. The 2004 image shows that much of that mid-latitude activity is subsiding, perhaps a return to a Voyager-era cloud distribution.

Fig. 2.—Uranus images with Keck 10-meter telescope. We imaged Uranus with the NIRC2 camera and adaptive optics system over several years. The south pole of Uranus is visible to the left. (a) At K' (2.2 $\mu$m), methane absorption darkens the planet, revealing the ring system and only those clouds that penetrate to relatively high altitudes (above the bulk of the absorbing methane). A yearly sequence (adapted from de Pater et al. 2005) shows the ring system closing as Uranus approaches equinox—i.e., ring plane crossing—in 2007 (left to right: 2001, 2002, 2003, 2004). (b) In this H (1.6 microns) image obtained on 9 July 2004, decreased methane absorption permits deeper layers to be imaged, revealing the prominent south polar collar and a significant amount of northern hemispheric activity. The bright ε ring is barely visible as a faint line crossing the planet.

Fig. 3.—Spectra of Neptune at 8-13 $\mu$m. Observations of Neptune in 2002 (yellow), 2003 (red), and 2004 (purple) are compared to data from 1991 (green, Hammel et al. 1992) and 1985 (blue, Orton et al. 1990, 1992). The 12.2-$\mu$m ethane emission brightened significantly from 1985 to 2003, returning to the 1991 level in 2004. The emission also changed shape during the past decade, indicating that another factor was operating in addition to temperature variations (temperature modifications scale only the amplitude of the emission). We see elevated methane emission near 8 $\mu$m in 2002-2004. Gray areas were used by Hammel et al. (1992) to calculate ethane/methane abundance ratios.

Fig. 4.—Neptune and Uranus atmospheric profiles. (a) We modeled Neptune's molecular emission using a pressure-temperature profile (solid) which is adapted from and consistent with earlier models (Orton et al. 1992; Moses et al., 1992); the Uranus profile (dotted) is from P. Romani (personal communication). (b) We examined two profiles for Neptune ethane mixing ratio as a function of pressure: the solid curve follows Orton et al. (1992); the dashed curve is represents Moses et al. (1992) as described in the text. Uranus ethane mixing ratio used in the models is shown as a dotted line (P. Romani, personal communication). For both figures, we plot only the pressure region relevant to the mid-infrared ethane molecular emission. (c) We show the monochromatic contribution functions for the peaks of two representative ethane lines (one strong, one weak) near 12.2 $\mu$m assuming a slight temperature gradient in the stratosphere [P. Romani, personal communication; c.f. Orton et al. (1992) in (a)]. The contribution functions for both lines peak near pressures of 0.03-0.05 mbar; an isothermal model also peaked near this atmospheric pressure range. A weak subsidiary peak near 1 mbar is from the continuum formed by $H_2$-dimer opacity.

Fig. 5.—BASS detector-element response function and Neptune's ethane emission. A typical response function of an individual detector element as measured in the lab (solid line) is plotted over a high-spectral-resolution model of Neptune's atmosphere within the 12-micron window.

Hammel, H. B., et al. 11Several lines of ethane contribute to the integrated spectral response within a detector element. If the detector-element response is shifted fractionally in wavelength (dotted line), then the specific ethane lines can shift from a peak (higher responsivity) to a valley (lower responsivity). The inset shows integrated detector-element flux as a function of detector-element offset (in microns). The two circled points correspond to the detector-element positions marked with solid and dotted lines in the main graph.

Fig. 6.—Effect of detector-element response variations. The non-uniformity described in the text and shown in Fig. 5 can lead to a variety of shapes for a single model of Neptune's atmosphere. Each plot corresponds to a constant wavenumber shift of the entire detector-element array relative to an assumed initial reference position taken as the 2002 calibration. The bottom curve corresponds to a 0 cm$^{-1}$ shift (i.e., the reference position); the other curves correspond to steps of -0.4 cm$^{-1}$ relative to the reference. Points are plotted in the bottom curve to illustrate that the ethane spectrum is sparsely sampled. The pattern repeats about every 1.6 cm$^{-1}$. Since a typical detector-element width corresponds to 7 cm$^{-1}$, all shapes can appear by invoking only a slight fraction of a detector-element shift in the wavelength positions from run to run over a period of years. The predicted shapes correspond reasonably well with the observed shapes; although the quantitative agreement is not perfect, and discrepancies probably reflect uncertainty in the exact detector-element response function.

Fig. 7.—Representative Neptune models. The thick colored lines represent typical RT models (Table 2) that reproduce the data (thin lines with symbols. reproduced from Fig. 3). The band's amplitude is most sensitive to stratospheric temperature changes; shape changes are likely due to the BASS detector-element spectral response as discussed in the text. Each data set has been adjusted in wavelength as listed in Table 2.

Fig. 8.—Long-term variation of Neptune. (a) Solid circles indicate $T_{eff}$ determined here. Asterisks are from Marten et al. (2005): their initial nominal model based on Voyager observations (date: 1989.72); the value used by Bézard et al. (1999) to model ISO observations (date: 1997.95); and the value Marten et al. deduce from sub-mm CO observations (date: 1998.75). The sub-mm observations provide an independent measure of temperature change in the atmosphere of Neptune, consistent with our results. (b) Long-term disk-integrated photometry of Neptune at Strömgren $y$ from Lockwood & Jerzykiewicz (2006).

Fig. 9.—Average Uranus spectrum from 11 to 13 $\mu$m, with representative model (gray) and its principal components: ethane (dark blue), acetylene (light blue) and the H$_2$ continuum emission (red). We integrated on Uranus for over 7 hrs (see Table 1) to produce the spectrum (points with error bars). The model is discussed in the text. The discrepancy near the peak emission is likely due to the detector-element response function, as discussed for Neptune. Shortward of 11.5 $\mu$m, we speculate that some other species (even though two orders of magnitude less in mixing ratio than ethane) could be contributing flux at this wavelength. However, the error bars are large and the data at the noise level. Spitzer observations may provide insight on this wavelength region, and thus we prefer to defer quantitative speculation.

Fig. 10.—Long-term variation of Uranus. (a) Solid circle indicates $T_{eff}$ determined here (the uncertainty of ±3 K is smaller than the symbol); asterisks are from Young et al. (2001). The occultation observations provide an independent measure of temperature change in the atmosphere of Uranus, consistent with our results and with seasonal variation. (b) Long-term disk-integrated photometry of Uranus at Strömgren $y$ from Lockwood & Jerzykiewicz (2006).

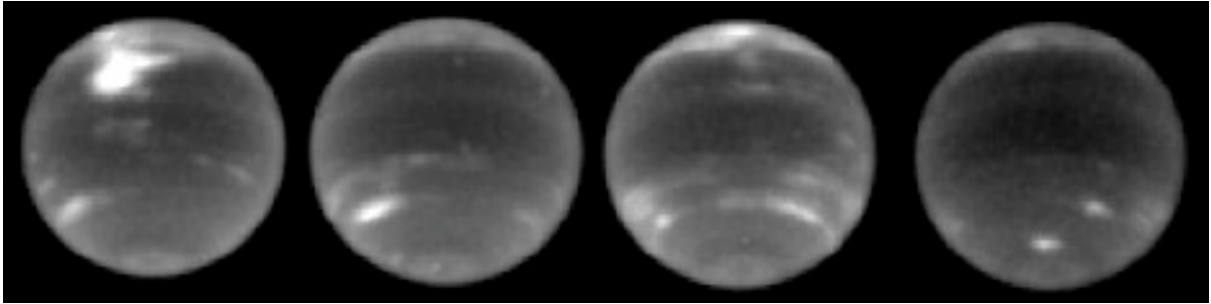

**FIGURE 1**

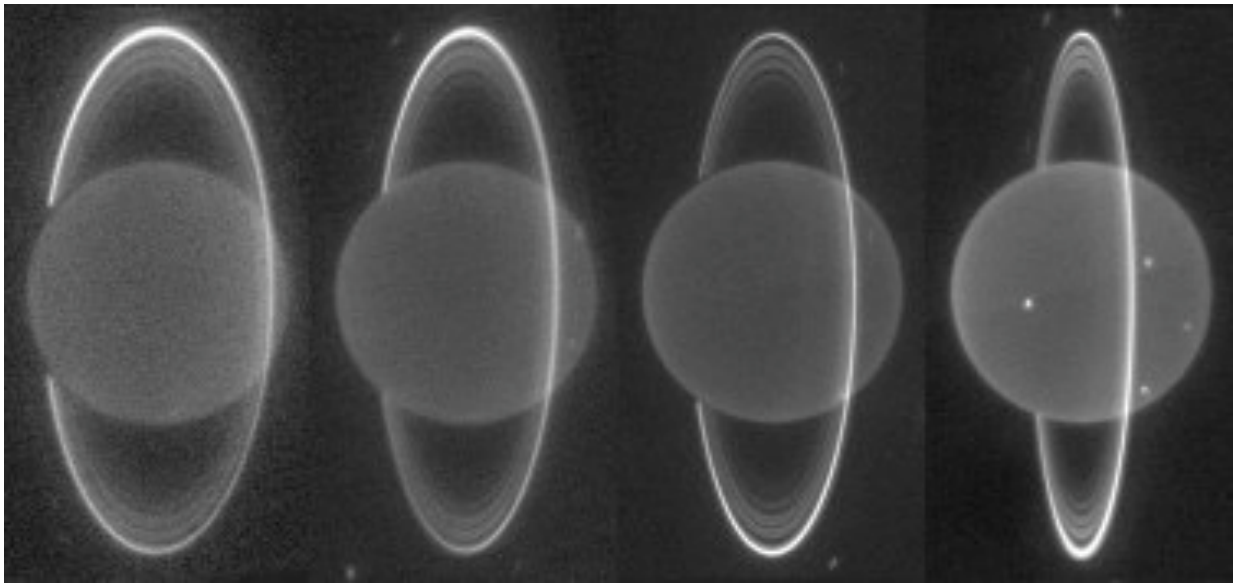

**FIGURE 2a**

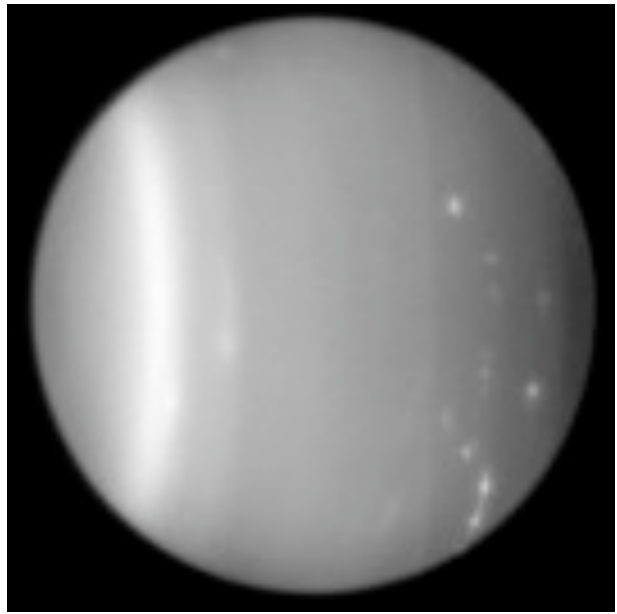

**FIGURE 2b**

Hammel, H. B., et al.     13

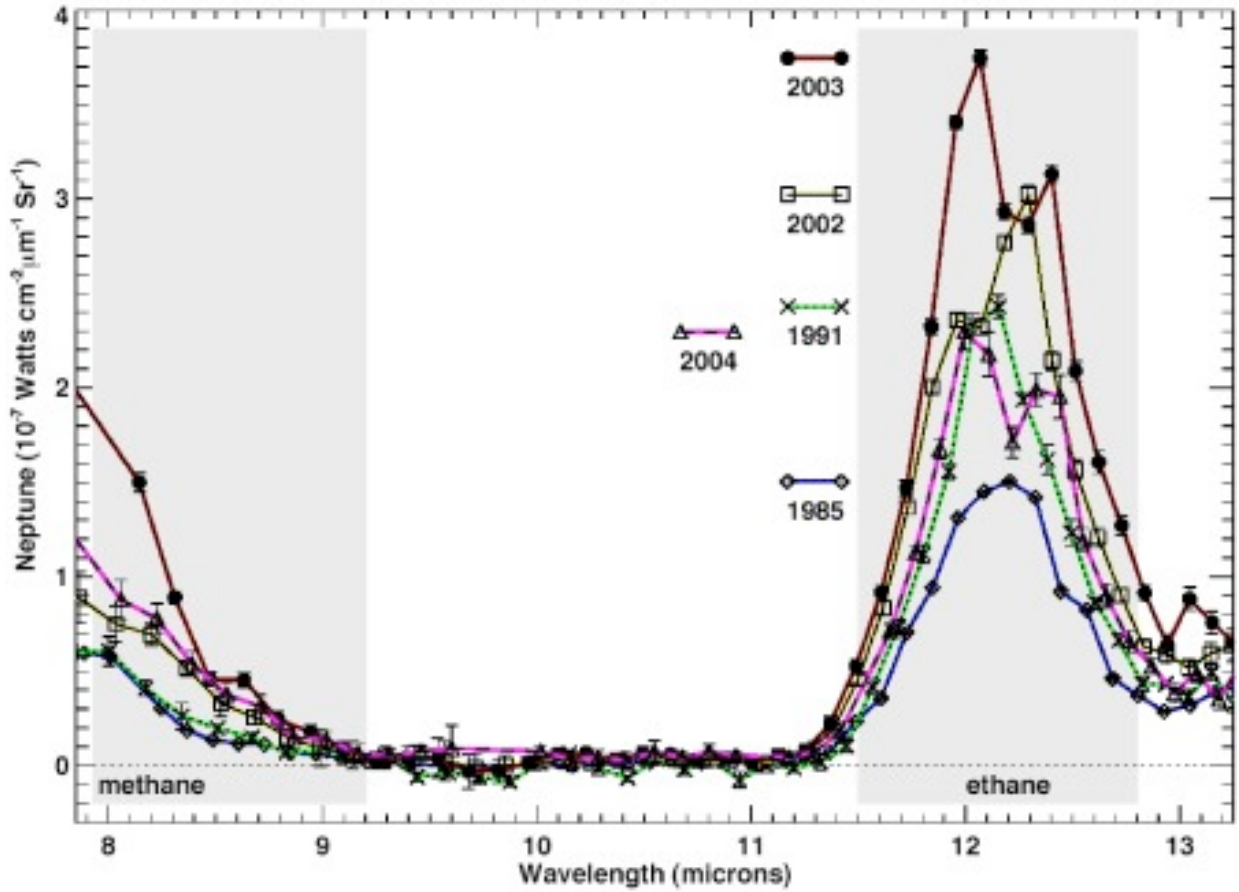

FIGURE 3

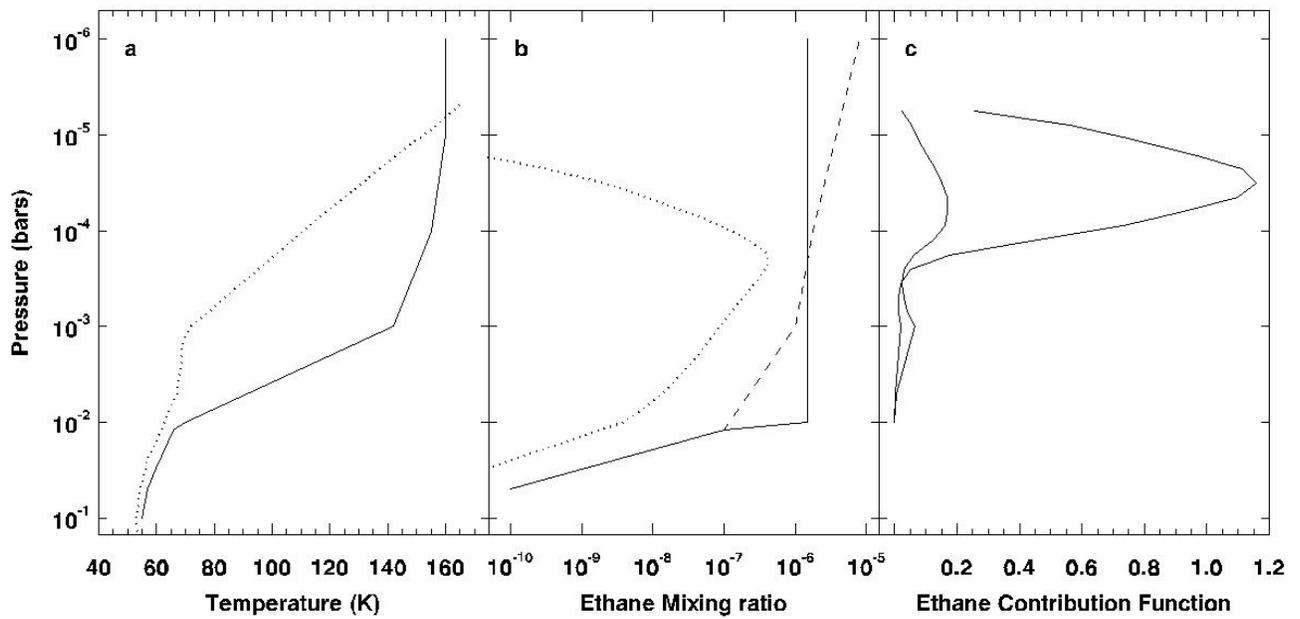

FIGURE 4

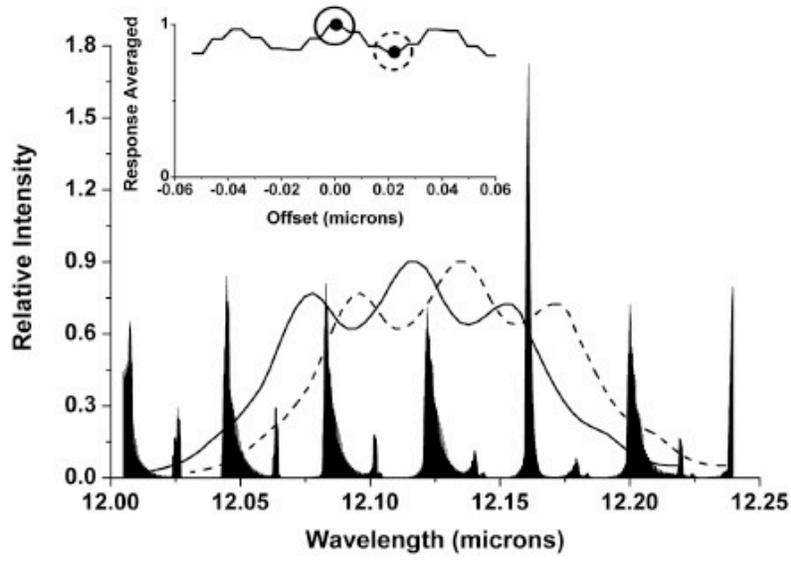

**FIGURE 5**

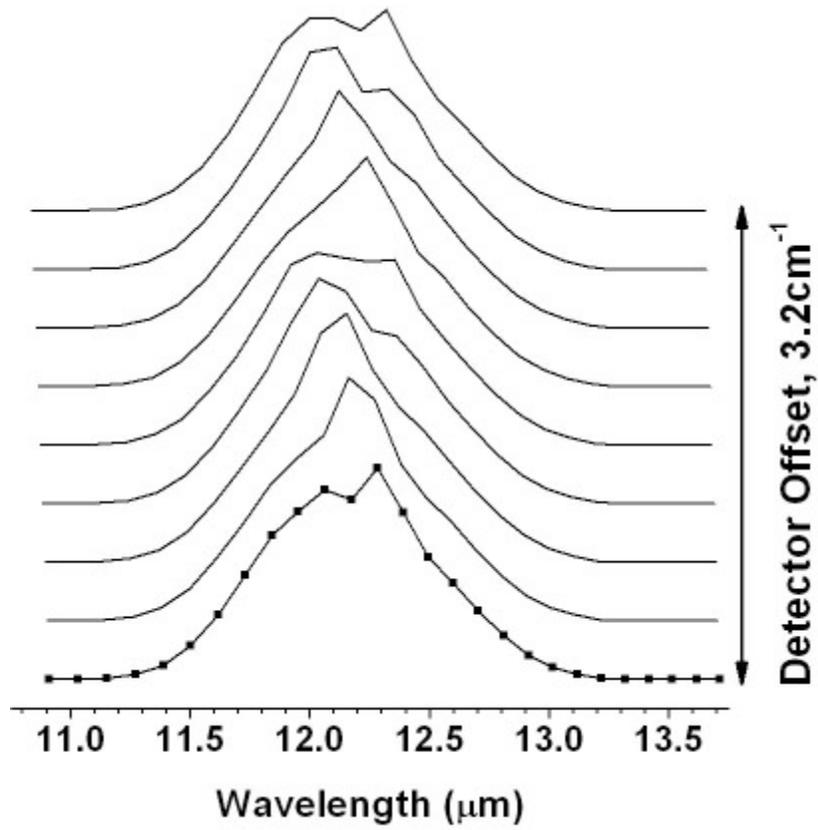

**FIGURE 6**

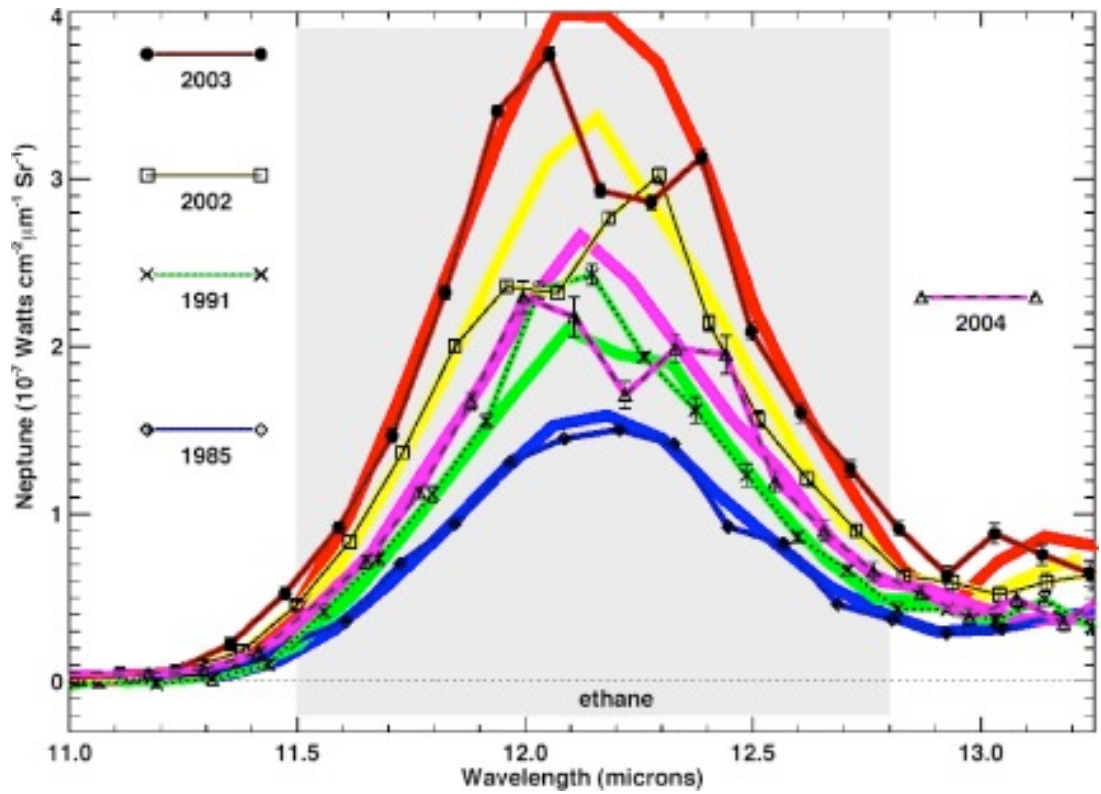

**FIGURE 7**

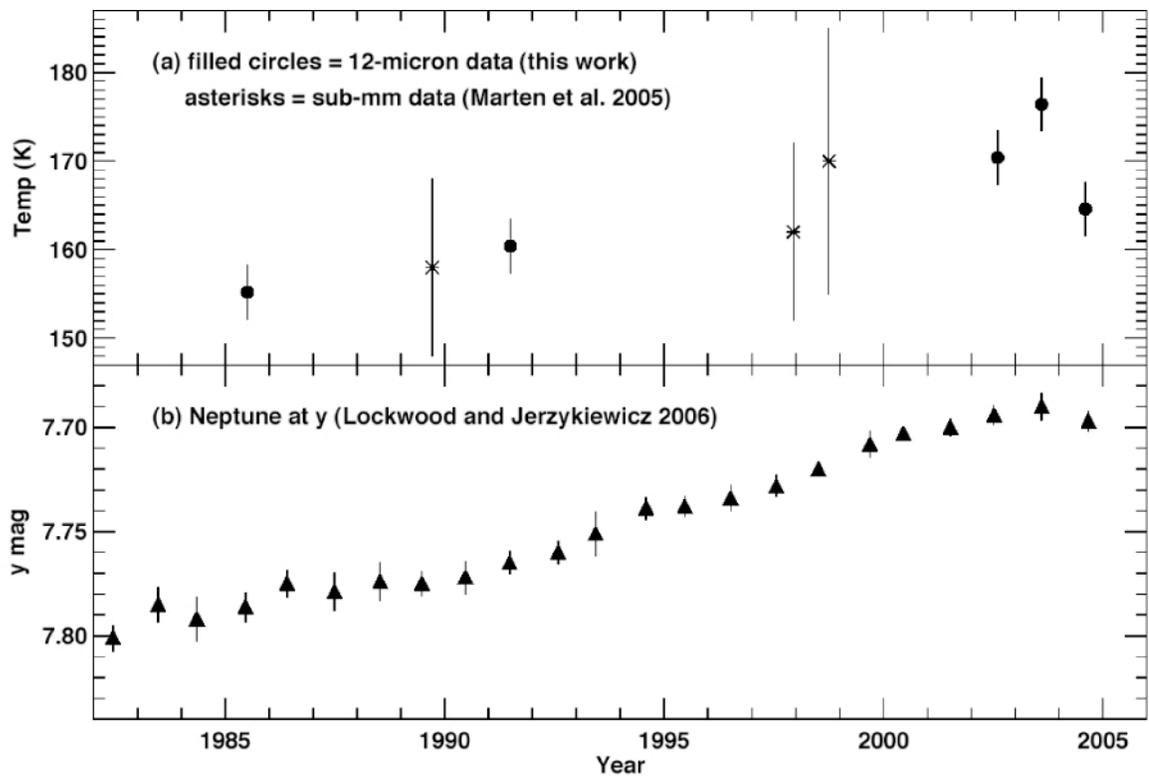

**FIGURE 8**

Ethane on Neptune and Uranus 16

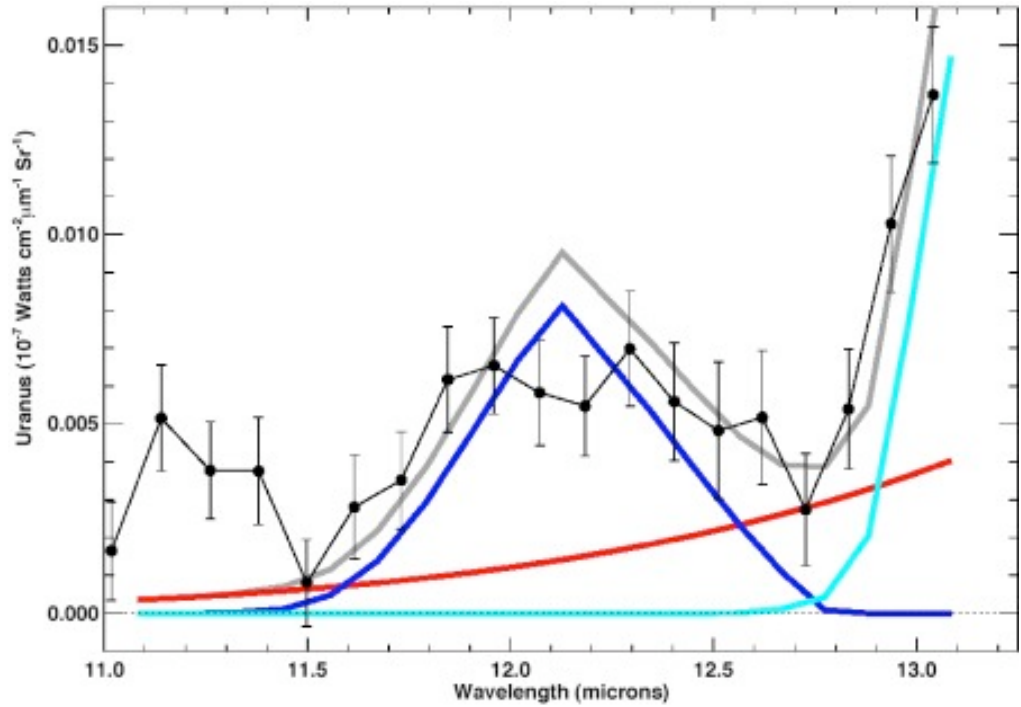

FIGURE 9

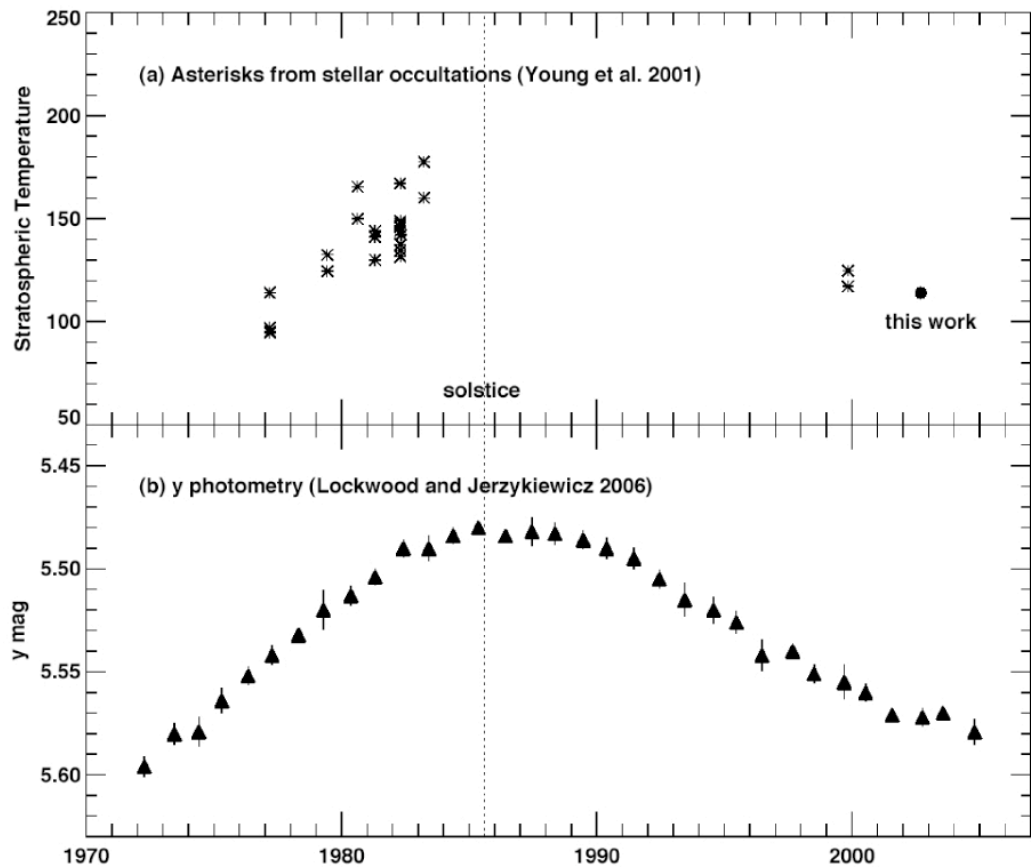

FIGURE 10